\documentclass[prd,aps,floats,twocolumn,nofootinbib]{revtex4-1}
\usepackage{slashed}
\usepackage{mathtools}
\usepackage{amsfonts}
\usepackage{amssymb}
\usepackage{amsmath}
\usepackage{comment}
\usepackage{xcolor}
\usepackage{epsfig}
\usepackage{physics}
\usepackage{epsfig}

\begin{document}

\newcommand{\m}[1]{\mathcal{#1}}
\newcommand{\nn}{\nonumber}
\newcommand{\ph}{\phantom}
\newcommand{\eps}{\epsilon}
\newcommand{\be}{\begin{equation}}
\newcommand{\ee}{\end{equation}}
\newcommand{\bea}{\begin{eqnarray}}
\newcommand{\eea}{\end{eqnarray}}
\newtheorem{conj}{Conjecture}

\newcommand{\plk}{\mathfrak{h}}
\newcommand{\bb}{\bar b}


\title{Spontaneous Symmetry Breaking in Graviweak Theory}
\date{\today}

%

\author{Stephon Alexander$^1$, Bruno Alexandre$^2$, Michael Fine$^2$, Jo\~{a}o Magueijo$^2$, Max Pezzelle$^1$}
\email{stephon\_alexander@brown.edu}
\email{bruno.alexandre20@imperial.ac.uk}
\email{j.magueijo@imperial.ac.uk}
\email{max\_pezzelle@brown.edu}

\affiliation{$^1$Department of Physics, Brown University, Providence, RI 02912, USA}
\affiliation{$^2$Abdus Salam Centre for Theoretical Physics, Imperial College London, Prince Consort Rd., London, SW7 2BZ, United Kingdom}


\begin{abstract}
Graviweak theory seeks to unify gravity (specifically in its self-dual formulation) with the weak interaction, preying on their parallel chiral $SU(2)$ structures. In this paper we further this idea by folding it with the concept of spontaneous symmetry breaking. We do this first with a standard Higgs field and potential, starting with a unifying parity-invariant theory which splits into the usual gravity and weak sector under spontaneous symmetry breaking. By rewriting the theory in the two-spin framework we are then prompted to discuss generalizations, within the generic approach known as MacDowell-Mansouri theories where a larger internal gauge group is broken. One of the predictions of the ensuing construction is a  non-minimal coupling in the low energy broken phase between curvature and the weak gauge fields, translating at the quantum level to a direct channel between the graviton and the weak bosons. 
\end{abstract}

\maketitle

\section{Introduction and Motivation}

The unification of gravity with the other fundamental interactions remains
one of the most profound challenges in theoretical physics. While the Standard Model successfully unifies the electroweak and strong interactions within
a gauge-theoretic framework, gravity remains conceptually and technically distinct due to its geometric nature.
An intriguing approach to unification is the idea that gravity and the weak
interaction could emerge from a common underlying chiral gauge theory. This
framework is referred to as {\it graviweak} unification \cite{Alexander:2012ge,Alexander:0706.4481,Nesti:0706.3307,Nesti:0706.3304}. The motivation stems from
the realization that both gravity and the weak interaction can be formulated as chiral theories. The
weak interaction is intrinsically chiral, only left-handed fermions couple to the
weak gauge bosons, while gravity, in the Einstein-Cartan formalism, can also
be made chiral by splitting the spin connection into left- and right-handed components associated with the group structure \cite{KirillBook,Freidel:2005mq,Capovilla:1991qt,Plebanski:1977pw,Celada:2016jdt,Alexandre:2407.19363}
\[
SO(4)\cong SU(2)_L\times SU(2)_R.
\]
The parity asymmetry of the weak interaction suggests that parity should
also be broken in the gravitational sector at a high energy scale. If the chiral asymmetry
of the weak interaction arises from spontaneous symmetry breaking \cite{Wilczek:1998ea}, a similar
mechanism could break parity in the gravitational sector, leaving only a left-chiral gravitational connection at low energies.

These considerations motivate the construction of a theory in which gravity is initially described by both chiral components. A spontaneous symmetry-breaking mechanism then suppresses the
right-handed chiral component at low energies. The left-handed component would naturally couple to left-handed fermions, mimicking the chiral nature of the weak interaction. In particular, the gravitational interaction should
have the opposite chirality of the weak interaction, reflecting the parity-breaking
structure of the Standard Model.
This paper develops a model of graviweak unification where a Higgs-like field \cite{higgs}
generates this parity asymmetry dynamically through a $Z_2$ symmetric potential.
The right-handed kinetic term is suppressed, leaving a chiral
gravitational sector.

We develop these ideas in this paper along the following logical thread. In Section~\ref{GWH} we propose a gravitational theory formulated in terms of chiral components of the Lorentz connection and tetrad fields. The total Lagrangian includes Einstein-Hilbert-type terms, Yang-Mills-like kinetic terms for the chiral curvatures, and a Higgs-like pseudoscalar field \( \phi \) that couples to the the parity odd terms and dynamically breaks parity symmetry through its vacuum expectation value. Thus, we have a theory that unifies gravity and the weak interaction in a parity invariant construction before symmetry breaking, reducing to the Ashtekar self-dual formulation of gravity plus the weak sector after symmetry breaking. This will be the foundations of this paper, around which we will investigate developments and variations. 

In Section~\ref{AA'GWH} we then rewrite this theory in the 
two-spinor notation, the natural language for treating gravity and chiral gauge fields on the same footing. This formalism is especially convenient for formulating unification frameworks such as graviweak models and it points us towards a new development: parity symmetry breaking that goes hand in hand with a larger gauge group. This receives further exploration in Section~\ref{GWHMM}, where we extend the internal Lorentz group underpinning the basic construction to $SO(1,3+N)$. A major implication is the appearance of possible interactions between the gravitational 2-form curvature $R^a_{~b}$ of $SO(1,3)$ and the Yang-Mills field strength $F^{ij}$ of a $SO(N)$ group. This is a remarkable novelty with regards to previous graviweak models. We close this paper with a few speculations on the testability and falsifiability of this scenario.

\section{The Graviweak-Higgs System}\label{GWH}
We start by laying down the structure of the theory, which is initially parity invariant and built from left $(+)$ and right-handed $(-)$ chiral curvature 2-forms. 
This is done by introducing the projection operators: 
\begin{eqnarray}
    P^{(\pm)ab}{}_{cd}=\frac{1}{2}\left(\delta^{ab}_{cd}\pm\frac{i}{2}\epsilon^{ab}{}_{cd}\right),
    \label{Projectors}
\end{eqnarray}
where the chiral components of the Cartan spin connection 
$\omega_{ab}$ are given by:
\begin{eqnarray}
     P^{(\pm){cd}}{}_{ab}\;\omega_{cd}=\omega^{(\pm)}_{ab} = \frac{1}{2} \left( \omega_{ab} \pm \frac{i}{2} \epsilon_{ab}{}^{cd} \omega_{cd} \right), \label{omega}
\end{eqnarray}
and hence the curvature 2-forms are:

\begin{equation}
    R^{(\pm)}_{ab} = d\omega^{(\pm)}_{ab} + \omega^{(\pm)}_a{}^c \wedge \omega^{(\pm)}_{cb}.
\end{equation}
We note the following useful definitions of Hodge stars. The $\widetilde {\ }$ denotes the Hodge dual on the spacetime indices and the star symbol $\star$ will be used to denote the Hodge dual on the group indices:
\begin{eqnarray}
    \widetilde R^{ab}_{\mathrm{\ \ }\mu\nu} = \frac{1}{2}\epsilon_{\rho\sigma\mu\nu}R^{ab\rho\sigma},\nonumber\\
     \star R^{ab}_{\mathrm{\ \ }\mu\nu} = \frac{1}{2}\epsilon^{abcd}R_{cd\mu\nu}.\nonumber
\end{eqnarray}
These definitions coupled with equation \eqref{omega}, show that applying the Hodge star in the group, to chiral terms yields:
\begin{equation}
    {\star}R^{(-)}_{ab}\;=\;i\,R^{(-)}_{ab},
    \label{star-}
\end{equation}
\begin{equation}
{\star}R^{(+)}_{ab}\;=\;-\,i\,R^{(+)}_{ab}.
    \label{star+}
\end{equation}

With these in mind we then consider the most general parity-invariant gauge theory in $SO(1,3)$ containing only linear and quadratic terms in the curvature, with the condition that the theory should result in general relativity and the weak sector after spontaneous symmetry breaking. Its Lagrangian is

\begin{widetext}
    \begin{align}
\mathcal{L}_{\text{total}} =\; & \frac{1}{16\pi G} \left[\, \, e^a \wedge e^b \wedge \star R_{ab} +i\phi \:\:e ^a\wedge e^b\wedge R_{ab} \right]\nonumber\nonumber \\
&- \frac{1}{8g^2} \left[R^{}_{ab} \wedge \widetilde  R^{ab}+i\phi \: R_{ab} \wedge \star\widetilde  R^{ab}\right] \nonumber \\
&- \frac{1}{2} d\phi \wedge \widetilde  d\phi - {V(\phi)}\widetilde 1. \label{Ltotal}
\end{align}
\end{widetext}

A few comments on this action are in order. First note that 
the parity transformations under which this theory is initially symmetric are defined as follows. The tetrad transforms as \( e^a(x^0,\vec{x}) \mapsto e^a(x^0,-\vec{x}) \), and the chiral components of the connection exchange under parity: \( \omega^{(+)}_{ab}(x^0,\vec{x}) \mapsto \omega^{(-)}_{ab}(x^0,-\vec{x}) \), and vice versa. Consequently, the curvatures transform as \( R^{(+)}_{ab}(x^0,\vec{x}) \mapsto R^{(-)}_{ab}(x^0,-\vec{x}) \) and \( R^{(-)}_{ab}(x^0,\vec{x}) \mapsto R^{(+)}_{ab}(x^0,-\vec{x}) \). The pseudoscalar field \( \phi(x^0,\vec{x}) \) is to be parity-odd, transforming as \( \phi(x^0,\vec{x}) \mapsto -\phi(x^0,-\vec{x}) \).

We consider a modification of the chiral gravitational action in which the couplings to parity odd terms such as the Holst term are modified to parity even by the pseudoscalar field \( \phi \).
This ensures that parity symmetry is respected at the level of the classical action. The Lagrangian is constructed to be explicitly parity-invariant, even while allowing for the spontaneous breaking of parity through the vacuum structure of \( \phi \): \( V(\phi) \) is a parity-invariant potential, of the Higgs form 
\[ V(\phi) = \frac{\lambda}{4} (\phi^2 - 1)^2, \] 
leading to spontaneous symmetry breaking with vacuum expectation values (vev) \( \langle \phi \rangle =  \pm1 \) and where \( \lambda \) is a coupling constant. Under a parity transformation, the exchange of the parity odd terms is exactly compensated by \( \phi \mapsto -\phi \), ensuring that the Lagrangian remains invariant.
Once \( \phi \) acquires a nonzero vacuum expectation value, this symmetry is spontaneously broken. The couplings induce a differential scaling between the self-dual and anti-self-dual sectors. For example, for \( \langle \phi \rangle > 0 \), one sees, once factoring out $i$ that the first line of the Lagrangian becomes the right handed projection operator in \eqref{Projectors} times a Holst term:  $\frac{i}{8\pi G}\;e^a \wedge e^b \wedge P^{(-)cd}{}_{ab}R^{cd} =\frac{i}{8\pi G}\;e^a \wedge e^b \wedge R^{(-)}_{ab}$; for \( \langle \phi \rangle < 0 \), the opposite occurs $\frac{i}{8\pi G}\;e^a \wedge e^b \wedge P^{(+)cd}{}_{ab}R^{cd} =\frac{i}{8\pi G}\;e^a \wedge e^b \wedge R^{(+)}_{ab}$. This leads to dynamical chirality selection, where the vacuum structure of the pseudoscalar field determines which chiral sector governs the low-energy gravitational dynamics.

The field \( \phi \) in this framework behaves as a chiral modulus: it continuously controls the binary weighting of curvature contributions from the two chiral sectors. This construction introduces a novel, geometrically natural framework in which parity remains a fundamental symmetry of the action, while chirality can be selected dynamically through spontaneous symmetry breaking. It defines a new class of moduli-driven parity-variant theories of gravity within a chiral, pseudoscalar, and parity-respecting structure.

This theory is thus parity symmetric at the level of the action, and explicitly so before spontaneous symmetry breaking. However, once the pseudoscalar field condensates, the coupling between \( \phi \) and the chiral curvatures breaks the parity symmetry spontaneously, leading to asymmetric effective dynamics in the gravitational sector.
Substituting the vev into (\ref{Ltotal}) and employing equations \eqref{Projectors} and \eqref{star-} yields:
\begin{eqnarray}
   \mathcal{L}=- \frac{1}{4g^2} R^{(+)}_{ab} \wedge \widetilde  R^{ab(+)}  + \frac{1}{8\pi G}\;e^a \wedge e^b \wedge \star R^{(-)}_{ab} . \nonumber
\end{eqnarray}
 As a result, the right-handed connection becomes purely gravitational, meaning it is only present in the Lagrangian by coupling to the tetrad, and only the left-handed kinetic term remains:
\begin{eqnarray}
    \mathcal{L}_{\text{YM}}^{\text{eff}}=  -\frac{1}{4g^2} R^{(+)}_{ab} \wedge \widetilde R^{ab(+)}.
\end{eqnarray}

The parity even coupling of the fermions is introduced through the general covariant derivative:

\begin{eqnarray}
\mathcal{L}_{\text{fermion}}=-\frac{i}{2}\star e_a\wedge\bar \psi\gamma^aD^{}\psi.
\end{eqnarray}

The spin connection can then be decomposed into chiral parts,  
and thus the exterior covariant derivative acting on the spinors is defined by
\begin{eqnarray}
D^{(\pm)}\psi_{L/R}=d\psi_{L/R}+\frac{1}{4}\omega^{(\pm)ab}\gamma_{ab}\psi_{L/R},
\end{eqnarray}
where $\psi_{L/R}=P_{L/R}\psi$ are the usual left- and right-handed spinor components and $\gamma_{ab} = \gamma_{[a}\gamma_{b]}$. It can be shown that the product of projection operators of the opposing chirality between $P_{L/R}$ and $P^{(\pm)}$ is zero and such the fermion Lagrangian reduces to: 
\begin{eqnarray}
\mathcal{L}_{\text{fermion}}=-\frac{i}{2}\star e_a\wedge[\bar \psi_L\gamma^aD^{(+)}\psi_L+\bar \psi_R\gamma^aD^{(-)}\psi_R].
\end{eqnarray}

Hence, after symmetry breaking and adding the spinor terms to (\ref{Ltotal}) we obtain:
\begin{align}
\mathcal{L}^{\text{eff}}_{\text{total}} =\; & \frac{1}{16\pi G} \, \epsilon^{abcd} \, e_a \wedge e_b \wedge R^{(-)}_{cd} \nonumber \\
&- \frac{1}{4g^2} \bar{F}_{i} \wedge \widetilde{\bar{F}}^{i} \nonumber \\
&-\frac{i}{2}\star e_a\wedge[\bar \psi_L\gamma^aD^{(+)}\psi_L+\bar \psi_R\gamma^aD^{(-)}\psi_R], \label{Ltotaleff}
\end{align}
which is of course standard gravity in the Ashtekar self-dual formulation plus weak theory, with fermions. We have relabeled the weak curvature as $\bar{F}^{i}=2iR^{(+)0i}$ to put it in a more familiar form.  Notice we also see that the weak sector only couples to left handed spinors, reflecting chirality and parity violation, and the right-handed ones only couple to gravity.

The equations of motion of our theory can be derived by the usual variation of the Lagrangian with respect to the fields, these being the tetrad $e^a$, both left- and right-handed connections $\omega^{ab(\pm)}$ and the spinors $\psi_{L/R}$. Hence, we get, respectively,
\begin{eqnarray}
    && \epsilon_{abcd}\Bigg[\frac{1}{8\pi G}e^b\wedge R^{cd(-)}+ \nonumber\\ &&+\frac{i}{4}\bar\psi_L\gamma^bD^{(+)}\psi_L \wedge e^c\wedge e^d + \frac{i}{4}\bar{\psi}_R \gamma^b D^{(-)}\psi_R \wedge e^c \wedge e^d \Bigg]=\nonumber\\
    &&=\frac{1}{4g^2}\left[R^{(+)cd}(\widetilde  R^{(+)}_{cd})_{ab}-(R^{(+)cd})_{ab}\widetilde  R^{(+)}_{cd}\right]e^b \nonumber \\
    && D^{(+)}\widetilde  R^{(+)}_{ab}=\frac{ig^2}{4}\star e_c \bar\psi_L\gamma^c \gamma_{ab}\psi_L \nonumber\\
    && \epsilon_{abcd}T^{(-)c}\wedge e^d= 2\pi Gi \star e_c\wedge \bar{\psi}_R\gamma^c \gamma_{ab}\psi_R\nonumber\\
    && \star e_a\gamma^a\wedge D^{(+)}\psi_L=0,\nonumber \\ && \star e_a\gamma^a\wedge D^{(-)}\psi_R=0,
\end{eqnarray}
where $T^{a(-)}=D^{(-)}e^a$ is the right-handed torsion of the tetrad.

In order to recover the Einstein-Cartan theory and $SU(2)$ Yang-Mills in Lorentzian signature, we will need to impose reality conditions \cite{Jacobson:1988fb}. These are the following:
\begin{equation}
    \omega^{(-)}_{ab} + (\omega^{(-)}_{ab})^* = \omega_{ab}^{(-)}(e)
\end{equation}
and
\begin{equation}
    \bar{A}_i = (\bar{A}_i)^*
\end{equation}
where $\omega^{(-)}_{ab}(e)$ is the right-handed spin connection which possesses vanishing torsion: $\mathrm{d}\omega^{(-)}_{ab}(e) + \omega^{(-)}_{ab}(e)\wedge e^b = 0$. Furthermore, $\bar{A}_i = 2i\omega^{(+)}_{0i}$ is the $SU(2)$ gauge field for $\bar{F}_i$:
\begin{equation}
    \bar{F}_i = d\bar{A}_i + \frac{1}{2}\epsilon_{ijk}\bar{A}^j \wedge\bar{A}^k.
\end{equation}

\section{Graviweak Lagrangian in Two-Spinor Notation}\label{AA'GWH}
The two-spinor notation is particularly adept for discussing parity issues so it is not surprising that it 
provides a natural language for treating gravity and chiral gauge fields on the same footing. This formalism is especially convenient for formulating unification frameworks such as graviweak models. As we will see towards the end of this Section, it leads to insights that justify the developments presented in the rest of this paper.

We begin by reviewing the two-spinor (or SL(2,$\mathbb{C}$)) notation.
In four-dimensional spacetime, the Lorentz algebra $\mathfrak{so}(3,1)$ is locally isomorphic to $\mathfrak{sl}(2,\mathbb{C}) \oplus \mathfrak{sl}(2,\mathbb{C})$, corresponding to the self-dual and anti-self-dual parts of the spin connection. Using the Infeld–van der Waerden map \cite{Jacobson:1988fb,Penrose:1984mf,Capovilla:1991kp}, spacetime indices $a = 0,1,2,3$ are replaced by spinor indices via the soldering form $\sigma^a_{AA'}$:
\begin{equation}
V^a \longleftrightarrow V^{AA'} = V^a \, \sigma_a^{AA'}.
\end{equation}
An antisymmetric tensor such as the spin connection $\omega^{ab}$ decomposes into self-dual $(-)$ and anti-self-dual $(+)$ parts as per (\ref{omega}), with:
\begin{equation}
\omega^{ab} = \omega^{(+)\,ab} + \omega^{(-)\,ab}.
\end{equation}
These map respectively to:
\begin{align}
\omega^{(-)\,ab} &\longleftrightarrow A^{AB}, \\
\omega^{(+)\,ab} &\longleftrightarrow \bar{A}^{A'B'},
\end{align}
where $A^{AB} = A^{(AB)}$ is the right-chiral (SD) SL(2,$\mathbb{C}$) connection, and $\bar{A}^{A'B'} = \bar{A}^{(A'B')}$ is its left-chiral (ASD) counterpart.
The tetrad $e^a$ is similarly rewritten as:
\begin{equation}
e^{AA'} = e^a \sigma_a^{AA'}.
\end{equation}
A detailed dictionary for the quantities appearing the last Section can then be developed and we refer the interested reader to Appendix~\ref{Appendix} for details.

We can therefore translate into this language our Lagrangian (\ref{Ltotaleff}). 
Using the spinor decomposition described above, we rewrite each term. The self-dual  Einstein-Cartan term becomes:
\begin{equation}
\mathcal{L}_{\text{grav}} = \frac{i}{16\pi G} \, e^{AA'} \wedge e^{BB'} \wedge R^{(-)}_{AA'BB'},
\end{equation}
where $R^{(-)}_{AA'BB'}=\epsilon_{A'B'}F_{AB}$ and $F^{AB}$ is the curvature 2-form associated with the self-dual connection $A^{AB}$. Therefore we have 
\begin{equation}
\mathcal{L}_{\text{grav}} = -\frac{i}{16\pi G} \, e^{AA'} \wedge e^{B}_{\,\,\,A'} \wedge F_{AB}.
\end{equation}
The curvature $\bar{F}^{A'B'}$ of the ASD connection contributes:
\begin{equation}
\mathcal{L}_{\text{YM}} = -\frac{1}{2g^2} \, \bar{F}^{A'B'} \wedge \widetilde {\bar{F}}_{A'B'},
\end{equation}
where $\bar{F}^{A'B'}$ is the field strength 2-form associated with $\bar{A}^{A'B'}$.

The fermion kinetic term involves the covariant derivative with respect to the ASD connection and the Dirac gamma matrix $\gamma^a$ is translated into the soldering form:

\begin{equation}
\mathcal{L}_{\text{fermion}} = -\frac{i}{2} \, \star e^{AA'} \wedge [\bar{\psi}_{A} \, D^{(+)} \psi_{A'}+\bar{\psi}_{A'} \, D^{(-)} \psi_{A}],
\end{equation}
where $D^{(\pm)}$ is the spinor covariant derivative acting on left- and right-handed spinors $\psi_{A'}$ and $\psi_{A}$ via the connections $\bar A^{A'}_{\,\,\,\,B'}$ and $A^{A}_{\,\,\,\,B}$. We also have
\begin{equation}
    \star e_{AA'} = \frac{i}{3}e_A^{\,\,\,\,C'}\wedge e_{DC'}\wedge e^D_{\,\,\,\,A'}.
\end{equation}
The full Lagrangian in 2-spinor form becomes:
\begin{widetext}
\begin{equation}
\mathcal{L} =
-\frac{i}{16\pi G} \, e^{AA'} \wedge e^{B}_{\,\,\,A'} \wedge F_{AB}
-\frac{1}{2g^2} \, \bar{F}^{A'B'} \wedge \widetilde{\bar{F}}_{A'B'}
- \frac{i}{2} \, \star e^{AA'} \wedge [\bar{\psi}_{A} \, D^{(+)} \psi_{A'}+\bar{\psi}_{A'} \, D^{(-)} \psi_{A}].
\end{equation}
\end{widetext}
This form makes manifest the chiral structure of the graviweak theory and sets the stage for further unification via larger gauge groups, as we now show.

\section{Extension of Lorentz Group -- $\boldsymbol{SO(1,3+N)}$}\label{GWHMM}

Bearing this motivation in mind, in this Section we extend the Lorentz group to $SO(1,3+N)$. A major implication will be the appearance of possible interactions between the gravitational 2-form curvature $R^a_{~b}$ of $SO(1,3)$ and the Yang-Mills field strength $F^{ij}$ of a $SO(N)$ group. This is a remarkable novelty with regards to previous graviweak models. 

\subsection{Structure of the Gauge Fields}

Let us start by considering the Lie algebra decomposition:
    \begin{eqnarray}
        \mathfrak{so}(1,3+N)\;\simeq\;\mathfrak{so}(1,3)\oplus\mathfrak{so}(N)\oplus(\text{coset})
    \end{eqnarray}
The generators $M_{AB}$ obey the commutation relation $
[M_{AB}, M_{CD}] = \eta_{AD} M_{BC} - \eta_{AC} M_{BD} - \eta_{BD} M_{AC} + \eta_{BC} M_{AD}
$
and have indices running as $A,B,C...=a,b,c,...,i,j,k,...=0,...,3+N$, where $a,b,c,..=0,1,2,3$ are in \( \mathfrak{so}(1,3) \) and $i,j,k,...=4,...,3+N$ in \( \mathfrak{so}(N) \). We therefore have $\frac{(N+4)(N+3)}{2}$ generators, which break into the 6 generators of $SO(1,3)$ plus the $\frac{N(N-1)}{2}$ generators of $SO(N)$, and the remaining ones are associated with the coset.

The gauge field is valued in the Lie algebra of \( \mathfrak{so}(1,3+N) \), with gauge field components \( \mathcal{A}^{AB} \) and its field strength is given by:
\[
\mathcal{F} = d\mathcal{A} + \mathcal{A} \wedge \mathcal{A}=\frac{1}{2}\mathcal{F}_{AB}M^{AB},
\]
where $\mathcal{A}=\omega+A+\phi$, $\omega=\frac{1}{2}\omega_{ab}M^{ab}$ is the spin connection, $A=\frac{1}{2}A_{ij}M^{ij}$ is the Yang-Mills connection and $\phi=\phi_{ai}M^{ai}$ the coset field.
We can write the total curvature in components as:
\begin{eqnarray}
    \mathcal{F}_{AB} = d\mathcal{A}_{AB} + \mathcal{A}_{AC} \wedge \mathcal{A}^C_{~B},
    \label{eqcalF}
\end{eqnarray}
and more explicitly:
\begin{eqnarray}
    && \mathcal{F}^a_{~b}=R^a_{~b}+\phi^a_{~i}\wedge \phi^i_{~b}, \\
    && \mathcal{F}^i_{~j}=F^i_{~j}+\phi^i_{~a}\wedge \phi^a_{~j}, \\
    && \mathcal{F}^a_{~i}=d\phi^a_{~i}+\omega^a_{~c}\wedge\phi^c_{~i}+\phi^a_{~j}\wedge A^j_{~i}.
\end{eqnarray}
The Killing form for this group is 
\[
\Tr(M_{AB}M_{CD})=2(2+N)(\eta_{AD}\eta_{BC}-\eta_{AC}\eta_{BD}).
\]
With these tools in hand we now consider a specific realization focusing on how it augments the construction in Sections~\ref{GWH} and \ref{AA'GWH}.

\subsection{The $\bold{SO(1,6)}$ case}

As an example let us consider the larger group  $SO(1,6)$. The total curvature is defined as in (\ref{eqcalF}) and an action quadratic in the curvature that we can consider in order to couple Yang-Mills and Einstein-Cartan is 
\begin{eqnarray}\label{so16quadraticpart}
    S &= \int \frac{\alpha}{2}\epsilon_{A_1...A_7}\mathcal{F}^{A_1A_2}\wedge\mathcal{F}^{A_3A_4}V^{A_5A_6A_7}\nonumber\\
    &+\int \alpha\mathcal{F}^{AB}\wedge \widetilde {\mathcal{F}}_{AB}+\frac{\lambda}{4}(V_{ABC}V^{ABC}-1)^2\widetilde 1,
\label{eqsi}
\end{eqnarray}
where we have introduced an $SO(1,6)$ Higgs field $V^{ABC}$ with the associated Mexican hat potential, and $\alpha$ and $\lambda$ are coupling constants.  In the unitary gauge we take a vev
\(V_0^{ijk}=\frac{1}{3!}\epsilon^{ijk}\)
(so \(\,V_0\cdot V_0=1\)),
which spontaneously breaks \(SO(1,6)\) into the internal $SO(3)$ block and the Lorentz $SO(1,3)$ block. Here
\(\epsilon^{ijk}\) is the Levi–Civita tensor of the internal \(SO(3)\). 

At this stage, one can consider breaking the symmetry in order to recover the Einstein-Cartan gravity sector along with the new interaction Einstein-Cartan/Yang-Mills interaction term. We follow a MacDowell-Mansouri-like symmetry breaking ansatz: 
\begin{eqnarray}
   && \phi^a_{~i}=\frac{1}{\ell}e^an_i , \,\,\,\, n_in^i=1, 
\end{eqnarray}
where $e^a$ is the $SO(1,3)$ 1-form tetrad, $n_i$ a normalized vector in $SO(3)$ and $\ell$ will be related to the cosmological constant $\Lambda$. Hence, the curvature terms become
\begin{eqnarray}
    && \mathcal{F}^a_{~b}=R^a_{~b}-\ell^{-2}e^a\wedge e_b, \label{fab}\\
    && \mathcal{F}^i_{~j}=F^i_{~j},\label{fij} \\
    && \mathcal{F}^a_{~i}=\frac{1}{\ell}\Bigl(
        T^{a}\,n_{i}
        + e^{a}\wedge P_i
     \Bigr),\label{fai}
\end{eqnarray}
where $P_i=dn^i+n_{j} A^{j}{}_{i}$, $T^{a} \equiv D e^{a} = de^{a} + \omega^{a}{}_{b} \wedge e^{b}$ is the torsion, and the first equation has the same form as the one in MacDowell-Mansouri theories \cite{MacDowell:1977jt}.
When the field $V$ condensates at its vev, the first term in (\ref{eqsi}) becomes 
\begin{eqnarray} \epsilon_{A_1...A_{7}}\mathcal{F}^{A_1 A_2} \wedge{\mathcal{F}}^{A_3 A_4} V^{A_5A_6A_7}=\epsilon_{abcd}\mathcal{F}^{ab}\wedge\mathcal{F}^{cd},\nonumber
\label{interaction1}
\end{eqnarray}
more specifically:
\begin{widetext}
\begin{eqnarray}
    \frac{1}{2} \epsilon_{abcd}\mathcal{F}^{ab}\wedge\mathcal{F}^{cd}=\frac{1}{2}\epsilon_{abcd}R^{ab}\wedge R^{cd}+\frac{1}{\ell^4}\Sigma^{ab}\wedge\widetilde{\Sigma}_{ab}-\frac{2}{\ell^2}R^{ab}\wedge \widetilde{\Sigma}_{ab},
\end{eqnarray}
\end{widetext}
where we defined the area $2-$form $\Sigma^{ab}=e^a\wedge e^b$.
The second term in (\ref{eqsi}) corresponds to the kinetic term of the gauge field with the spacetime hodge dual and it can be written explicitly in terms of its \(\mathfrak{so}(1,3) \) and \(\mathfrak{so}(3)\) components as
\begin{eqnarray}
&&\mathcal{F}^{AB} \wedge \widetilde{\mathcal{F}}_{AB}=\mathcal{F}^{ab} \wedge \widetilde{\mathcal{F}}_{ab}+2\mathcal{F}^{ai} \wedge \widetilde{\mathcal{F}}_{ai}+\nonumber\\
&&+\mathcal{F}^{ij} \wedge \widetilde{\mathcal{F}}_{ij}.\label{eqfsf}
\end{eqnarray}
Using equations (\ref{fab}) to (\ref{fai}), the terms above can be explicitly written as 
\begin{eqnarray}
    \mathcal{F}^{ab} \wedge \widetilde{\mathcal{F}}_{ab}=R^{ab} \wedge \widetilde{R}_{ab}+\frac{1}{\ell^4}\Sigma^{ab}\wedge \widetilde{\Sigma}_{ab}-\frac{2}{\ell^2}R^{ab}\wedge \widetilde{\Sigma}_{ab},\nonumber
\end{eqnarray}
and
\begin{eqnarray}
    \mathcal{F}^{ai} \wedge \widetilde{\mathcal{F}}_{ai}=\frac{1}{\ell^2}T^a\wedge\widetilde{T}_a+ \frac{3}{\ell^2}P^i\wedge\widetilde{P}_i,
\end{eqnarray}
where we use the fact that $n_i P^i = 0$.

The expansion around the vev, $V=V_0+\delta V$, leads to very interesting interactions between gravity and the weak force, but their detailed analysis is beyond the scope of this work. These terms are of the form
\begin{eqnarray}
&\epsilon_{a b c d\,i j k}\;
(\mathcal{F}^{a b}\wedge \mathcal{F}^{c d})\;
\delta V^{i j k}, \\
&\epsilon_{a b c d\,i j k}\;
\bigl(\mathcal{F}^{a b}\wedge \mathcal{F}^{c i}\bigr)\;
\delta V^{d j k}, \\
&\epsilon_{a b c d\,i j k}\;
\bigl(\mathcal{F}^{a b}\wedge \mathcal{F}^{i j}\bigr)\;
\delta V^{c d\,k}, \\
&\epsilon_{a b c d\,i j k}\;
\bigl(\mathcal{F}^{a i}\wedge \mathcal{F}^{b j}\bigr)\;
\delta V^{c d\,k}, \\
&\epsilon_{a b c d\,i j k}\;
\bigl(\mathcal{F}^{a i}\wedge \mathcal{F}^{j k}\bigr)\;
\delta V^{b c d}.
\end{eqnarray}
Among these, the term of utmost interest is the direct coupling between the gravitational curvature and the YM field strength of the weak force: 
\begin{eqnarray}
\mathcal{L}_{\text{RF}}=\alpha\epsilon_{a b c d}(R^{ab}-\ell^{-2}e^a\wedge e^b)\wedge \epsilon_{ijk}F^{i j}
\delta V^{c d\,k}.
\end{eqnarray}

The first two terms in the action (\ref{eqsi}) can then combine to give
\begin{widetext}
\begin{eqnarray}
    \frac{1}{2} \epsilon_{abcd}\mathcal{F}^{ab}\wedge\mathcal{F}^{cd}+\mathcal{F}^{ab} \wedge \widetilde{\mathcal{F}}_{ab}&=&\frac{1}{2}\epsilon_{abcd}R^{ab}\wedge R^{cd}+R^{ab} \wedge \widetilde{R}_{ab}+\frac{2}{\ell^4}\Sigma^{ab}\wedge\widetilde{\Sigma}_{ab}-\frac{4}{\ell^2}R^{ab}\wedge \widetilde{\Sigma}_{ab}\nonumber\\
    &&+F^{ij} \wedge \widetilde{F}_{ij}+\frac{1}{\ell^2}T^a\wedge\widetilde{T}_a+ \frac{3}{\ell^2}P^i\wedge\widetilde{P}_i.
\end{eqnarray}
\end{widetext}
Choosing the constants $\alpha$ and $\ell$ so that we can recover EC with cosmological constant $\Lambda$:
\begin{eqnarray}
\alpha &=& -\frac{1}{64\pi G}\,\frac{3}{\Lambda}\,,\qquad
\frac{1}{\ell^{2}} \;=\; \frac{\Lambda}{3}\,,
\label{constants}
\end{eqnarray} 
leads to the final Lagrangian written as:
\begin{widetext}
\begin{eqnarray}
    \mathcal{L}&=&\frac{1}{32\pi G}\epsilon_{abcd}\left(R^{ab}-\frac{\Lambda}{6}e^a\wedge e^b\right)\wedge e^c\wedge e^d-\frac{1}{64\pi G}\left(T^a\wedge\widetilde{T}_a+3P^i\wedge\widetilde{P}_i\right)\nonumber\\
    &&-\frac{3}{64\pi G\Lambda}\left(F^{ij} \wedge \widetilde{F}_{ij}+R^{ab} \wedge \widetilde{R}_{ab}+\frac{1}{2}\epsilon_{abcd}R^{ab}\wedge R^{cd}\right)+\mathcal{L}_{\text{RF}},
\end{eqnarray}
\end{widetext}
and it describes graviweak theory. This Lagrangian possesses nonminimal couplings between the curvature form and the Yang-Mills  fields \cite{Essig:2013lka}.

\section{Conclusions and Outlook}
In this paper we furthered the graviweak approach~\cite{Alexander:0706.4481,Alexander:2012ge} in two ways. First, we wedded it with the Higgs mechanism~\cite{higgs}, explaining how a graviweak construction could arise from a parity invariant action, with the symmetry broken by a Higgs field with a $Z_2$ symmetric potential. Second, we linked the ensuing construction with ideas inspired by \cite{MacDowell:1977jt}, and investigated whether the $SU(2)$ employed in graviweak could be embedded in a larger group, which is then spontaneously broken. The first development allows for graviweak theory to be understood within a framework ubiquitous in unification theories: spontaneous symmetry breaking. The second is motivated mainly by the drive to seek experimental predictions which might prove or disprove this theory. We found that in most constructions arising from larger groups a non-minimal coupling between the curvature and the Yang-Mills  fields is inevitable. At a quantum level this would imply a vertex involving the graviton and the weak gauge bosons. 

Obviously this is but a proof of concept for testability and falsifiability. Still, one wonders if the possible production of a graviton background in the early Universe  might not get modified by the presence of the weak sector. Parity violations in the graviton background would have dramatic effects on the CMB polarization \cite{Contaldi:2008iw,Lue:1999rpc,Saito:0705.3701}. With a new generation of gravitational wave detectors and CMB experiments expected to come online in the next decade, graviweak could be put to the test. 

Much work remains to be done. 
Further studies could explore detailed phenomenological models to quantify precisely how graviweak-induced parity violations could manifest observationally. 
The remarkable coincidence between the frequency sensitivity window of the upcoming LISA mission \cite{LISA:2024hlh} and the electroweak energy scale (see for example~\cite{Caprini:2018mtu}) has not escaped us. A non-minimal coupling between gravitons and the weak gauge bosons would have an impact on the stochastic gravitational wave background. Could our model induce a resonance in the production of gravity waves at the electroweak scale? And would its parity violating features make the corresponding spike inimitable via other mechanisms? 
Additionally, theoretical investigations could seek a deeper understanding of the interplay between graviweak symmetry breaking and established cosmological models, particularly in inflationary and pre-inflationary epochs. Exploring connections with quantum gravity approaches such as loop quantum gravity \cite{Rovelli:2004tv,Alexander:2011qfa} might also provide valuable insights into the fundamental structure of space-time and particle interactions. A variation of this model \cite{followup}, for example, would connect directly  with the work on area metrics emerging from spin foams \cite{Borissova:2022clg,Borissova:2023yxs}.

\acknowledgments
We thank Chiara Caprini, Aaron Hui, and Jacob Kuntzleman for helpful conversations. BA was supported by FCT Grant No. 2021.05694.BD and JM was partly supported by STFC Consolidated Grant ST/T000791/1.

\clearpage
\onecolumngrid
\appendix

\section{Translation into the two-spinor notation}\label{Appendix}

Consider the self- and anti-self-dual decomposition of all objects as in equation (\ref{omega}). We want to write the gravity sector in terms of its self- and anti-self-dual parts in both representations. The area form is given in terms of the tetra by $\Sigma^{ab}=e^a\wedge 
e^b $. Therefore, we obtain for the Einstein-Cartan sector:
\begin{eqnarray}
    &&\frac{1}{2}\epsilon_{abcd}\Sigma^{ab}\wedge R^{cd}=i(\Sigma^{(-)ab}\wedge R^{(-)}_{ab}-\Sigma^{(+)ab}\wedge R^{(+)}_{ab}). \label{eqSRab}
\end{eqnarray}
We use the Infeld–van der Waerden map, where spacetime indices $a = 0,1,2,3$ are replaced by spinor indices via the soldering form $\sigma^a_{AA'}$ and consider the following relations:
\begin{eqnarray}
    && \epsilon_{AA'BB'CC'DD'}=-i(\epsilon_{AC}\epsilon_{BD}\epsilon_{A'D'}\epsilon_{B'C'}-\epsilon_{AD}\epsilon_{BC}\epsilon_{A'C'}\epsilon_{B'D'}) \\
    && \epsilon^{AB}\epsilon_{BC}=\epsilon^A_{~C}=-\delta^A_C \\
    && \epsilon^{AB}\epsilon_{CB}=\epsilon_C^{~A}=\delta^A_C.
\end{eqnarray}
We also have the following decompositions, again in terms of self- and anti-self-dual components:
\begin{eqnarray}
    && R^{CC'DD'}=R^{(-)CC'DD'}+R^{(+)CC'DD'}=\epsilon^{C'D'}F^{CD}+\epsilon^{CD}\bar{F}^{C'D'} \\
    &&\Sigma^{CC'DD'}=\Sigma^{(-)CC'DD'}+\Sigma^{(+)CC'DD'}=\epsilon^{C'D'}\Sigma^{CD}+\epsilon^{CD}\bar{\Sigma}^{C'D'}, \label{eqsigma}
\end{eqnarray}
and we write $\Sigma^{CC'DD'}=e^{CC'}\wedge e^{DD'}$.
The mapping introduced above should yield
\begin{align}
\frac{1}{2}\epsilon_{abcd}\Sigma^{ab}\wedge R^{cd}=\frac{1}{2}\epsilon_{AA'BB'CC'DD'}\Sigma^{AA'BB'}\wedge R^{CC'DD'} ,
\end{align}
where the left hand side was computed above and the right hand side is, after several simplifications using the $\epsilon$ identities,
\begin{eqnarray}
\frac{1}{2}\epsilon_{AA'BB'CC'DD'}\Sigma^{AA'BB'}\wedge R^{CC'DD'}=2i(\Sigma^{AB}\wedge F_{AB}-\bar{\Sigma}^{A'B'}\wedge\bar{F}_{A'B'}). \label{eqSRAB}
\end{eqnarray}
Comparing this with (\ref{eqSRab}) we arrive at
\begin{eqnarray}
i(\Sigma^{(-)ab}R_{(-)ab}-\Sigma^{(+)ab}R_{(+)ab})=2i(\Sigma^{AB}F_{AB}-\bar\Sigma^{A'B'}\bar F_{A'B'}).
\end{eqnarray}
From (\ref{eqsigma}) we get
\begin{eqnarray}
    && \Sigma^{AB}=-\frac{1}{2}e^{AC'}\wedge e^B_{\,\,\,\,C'} \\
    && \bar\Sigma^{A'B'}=-\frac{1}{2}e^{CA'}\wedge e_C^{\,\,\,\,B'}
\end{eqnarray}
and so the SD part of the action can be written in terms of the tetrad components as
\begin{eqnarray}
    \mathcal{L}_{\text{GR}} =\frac{i}{8\pi G} \, \Sigma^{AB} \wedge F_{AB} =-\frac{i}{16\pi G} \, e^{AA'} \wedge e^{B}_{\,\,\,\,A'} \wedge F_{AB}. \nonumber
\end{eqnarray}
The kinetic Yang-Mills term becomes:
\begin{eqnarray}
     \mathcal{L}_{\text{YM}}=-\frac{1}{4g^2} R^{(+)\,ab} \wedge \widetilde  R^{(+)}_{ab}=-\frac{1}{2g^2}\bar{F}^{A'B'} \wedge \widetilde {\bar{F}}_{A'B'}.
\end{eqnarray}
The fermionic Lagrangian can be written explicitly in the spinor indices as
\begin{eqnarray}
   && \mathcal{L}_{\text{fermion}} = -\frac{i}{2} \star e^a \wedge \bar{\psi}_L \gamma_a D^{(+)} \psi_L=-\frac{i}{2}\star e_a\wedge\bar\psi_{A}\sigma^{a\,AA'}D^{(+)}\psi_{A'} \\
   && = -\frac{i}{2} \, \star e^{AA'} \wedge \bar{\psi}_{A} \, D^{(+)} \psi_{A'}
\end{eqnarray}
where
\begin{eqnarray}
    && D^{(+)} \psi_{A'}=d\psi_{A'}+\frac{1}{4}\bar A_{A'}^{\,\,\,\,B'}\psi_{B'}, \\
    && \bar A_{A'}^{\,\,\,\,B'}\psi_{B'}=\omega^{(+)ab}\sigma_{a\,CA'}\sigma_b^{CB'}\psi_{B'}
\end{eqnarray}
and
\begin{eqnarray}
    && \star e_a \sigma^{a}_{AA'}=\frac{1}{3!}\epsilon_{abcd}\wedge e^b\wedge e^c\wedge e^d ~\sigma^{a}_{AA'}=\frac{i}{3}e_A^{\,\,\,\,C'}\wedge e_{DC'}\wedge e^D_{\,\,\,\,A'}\equiv\star e_{AA'}. \nonumber
\end{eqnarray}

\section{}
An alternative action that accomplishes the spontaneous symmetry breaking needed to remove the left-handed Einstein-Hilbert term and the right-handed quadratic term is the following:
\begin{eqnarray}
S &=& \frac{\alpha}{2}\int \mathcal{F}^{ab}\wedge \widetilde{\mathcal{F}}_{ab}
+\frac{i\alpha}{4}\int \frac{\phi}{v}\,\epsilon_{abcd}\,\mathcal{F}^{ab}\wedge \widetilde{\mathcal{F}}^{cd}
+\nonumber\\
&& +i\alpha\int \frac{\phi}{v}\,\mathcal{F}^{ab}\wedge \mathcal{F}_{ab}
+\int V(\phi)\,\widetilde{1}\,.
\label{eqso13}
\end{eqnarray}

Here $\alpha$ is a coupling constant, $\phi$ is the pseudoscalar Higgs field and $V(\phi) = \frac{\lambda}{4} (\phi^2 - v^2)^2$ is the potential given above in section \ref{GWH}. This action is also parity invariant before symmetry breaking. The first term is a kinetic term for the curvature, the second is an $SO(1,3)$ invariant term combining both Hodge duals in spacetime and in group indices, and the third is a Pontryagin-like term. 
Upon symmetry breaking of the pseudoscalar, we can simplify this new action as follows. Further using (\ref{fab}) leads to 

\begin{eqnarray}
    && \frac{1}{2}\mathcal{F}^{ab} \wedge \widetilde{\mathcal{F}}_{ab}+\frac{i}{4}\epsilon_{abcd}\mathcal{F}^{ab}\wedge\widetilde{\mathcal{F}}^{cd}=\mathcal{F}^{(+)ab} \wedge \widetilde{\mathcal{F}}^{(+)}_{ab}=\nonumber \\
    && =R^{(+)ab} \wedge \widetilde{R}^{(+)}_{ab}-\frac{2}{\ell^{2}}\, \Sigma^{(+)}_{ab} \wedge \widetilde{R}^{(+)ab}+\frac{1}{\ell^{4}} \Sigma^{(+)}_{ab}\,  \wedge\widetilde{\Sigma}^{(+)ab}\label{eqf+},
\end{eqnarray}


Putting these together, equation (\ref{eqf+}) takes the form
\begin{eqnarray}
    \frac{1}{2}\mathcal{F}^{ab} \wedge \widetilde{\mathcal{F}}_{ab}+\frac{i}{4}\epsilon_{abcd}\mathcal{F}^{ab}\wedge\widetilde{\mathcal{F}}^{cd}&=&R^{(+)ab} \wedge \widetilde{R}^{(+)}_{ab}+\frac{i}{\ell^4}\Sigma^{(-)ab}\wedge\Sigma^{(-)}{}_{ab}+\frac{2i}{\ell^2}R^{(+)ab}\wedge \Sigma^{(+)}_{ab}
    \label{fpsfp}
\end{eqnarray}

Let us expand each term in the above equation in terms of its constituents as per the MacDowell-Mansouri formalism.
\begin{eqnarray}
    && \mathcal{F}^{ab} \wedge \mathcal{F}_{ab}=R^{ab} \wedge R_{ab}-\frac{2}{\ell^{2}}\, e_{a} \wedge e_{b} \wedge R^{ab}= \nonumber \\
    && =R^{ab} \wedge R_{ab}-\frac{2}{\ell^{2}}R^{(+)ab}\wedge \Sigma^{(+)}_{ab}-\frac{2}{\ell^{2}}R^{(-)ab}\wedge \Sigma^{(-)}_{ab}.\nonumber
\end{eqnarray}

After spontaneous symmetry breaking and the pseudoscalar field takes the value $\phi = v$, then the Lagrangian takes the form
\begin{eqnarray}
\nonumber\mathcal{L}&=&\frac{-2i \alpha }{\ell^{2}} \Sigma_{ab}^{(-)} \wedge R^{(-)ab}+\frac{i\alpha }{\ell^{4}}\Sigma^{(-)ab} \wedge \Sigma^{(-)}_{ab}+\\&&+{\alpha}R_{ab}^{(+)}\wedge \widetilde{R}^{(+)ab}+i\alpha R^{ab} \wedge R_{ab}.
\label{fulldecomposition}
 \end{eqnarray}
If we set
\begin{eqnarray}
\alpha &=& -\frac{1}{32\pi G}\,\frac{3}{2\Lambda}\,,\qquad
\frac{1}{\ell^{2}} \;=\; \frac{2\Lambda}{3}\, ,
\label{constants}
\end{eqnarray} 
then the Lagrangian takes the form similar to that shown in section \ref{GWH}:
\begin{eqnarray}
\nonumber\mathcal{L}&=&\frac{1}{32\pi G} \epsilon_{abcd}\left(e^a\wedge e^b \wedge R^{(-)cd}-\frac{\Lambda}{6}e^a\wedge e^b \wedge e^c\wedge e^d  \right)+\\&&-\frac{3}{64\pi G\Lambda} R_{ab}^{(+)}\wedge \widetilde{R}^{(+)ab}.
\label{fulldecomposition2}
\end{eqnarray}
Note that the gauge coupling in this model is controlled by the value of the gravitational constant and the cosmological constant.

\clearpage
\twocolumngrid


\begin{thebibliography}{99}

\bibitem{Alexander:2012ge}
S.~Alexander, A.~Marciano and L.~Smolin,
Phys. Rev. D \textbf{89}, no.6, 065017 (2014)
doi:10.1103/PhysRevD.89.065017
[arXiv:1212.5246 [hep-th]].

\bibitem{Alexander:0706.4481}
S.~H.~S.~Alexander,
``Isogravity: Toward an Electroweak and Gravitational Unification,''
[arXiv:0706.4481 [hep-th]].

\bibitem{Nesti:0706.3307}
F.~Nesti and R.~Percacci,
J.\ Phys.\ A \textbf{41}, 075405 (2008)
doi:10.1088/1751-8113/41/7/075405
[arXiv:0706.3307 [hep-th]].

\bibitem{Nesti:0706.3304}
F.~Nesti,
Eur.\ Phys.\ J.\ C \textbf{59}, 723 (2009)
doi:10.1140/epjc/s10052-008-0808-y
[arXiv:0706.3304 [hep-th]].

\bibitem{KirillBook}
K.~Krasnov, {\em Formulations of General Relativity: Gravity, Spinors and Differential Forms} (Cambridge University Press, 2020).

\bibitem{Freidel:2005mq}
L.~Freidel, D.~Minic and T.~Takeuchi,
Phys.\ Rev.\ D \textbf{72}, 104002 (2005)
doi:10.1103/PhysRevD.72.104002
[arXiv:hep-th/0507253].

\bibitem{Capovilla:1991qt}
R.~Capovilla, T.~Jacobson and J.~Dell,
Class.\ Quant.\ Grav.\ \textbf{8}, 59 (1991)
doi:10.1088/0264-9381/8/1/010

\bibitem{Plebanski:1977pw}
J.~F.~Plebanski,
J.\ Math.\ Phys.\ \textbf{18}, 2511 (1977).

\bibitem{Celada:2016jdt}
M.~Celada, D.~Gonz\'alez and M.~Montesinos,
Class. Quant. Grav. \textbf{33}, no.21, 213001 (2016)
doi:10.1088/0264-9381/33/21/213001
[arXiv:1610.02020 [gr-qc]].

\bibitem{Alexandre:2407.19363}
B.~Alexandre,
arXiv:2407.19363 [gr-qc].

\bibitem{Wilczek:1998ea}
F.~Wilczek,
Phys. Rev. Lett. \textbf{80}, 4851-4854 (1998)
doi:10.1103/PhysRevLett.80.4851
[arXiv:hep-th/9801184 [hep-th]].

\bibitem{higgs}
P. Higgs (1964), {\em Phys. Lett.} \textbf{12} (1964), 132; {\em Phys. Rev. Lett.} \textbf{13} (1964), 508

\bibitem{Jacobson:1988fb}
T.~Jacobson,
``Fermions in Canonical Gravity,''
Class.\ Quant.\ Grav.\ \textbf{5} (1988) L143.

\bibitem{Penrose:1984mf}
R.~Penrose and W.~Rindler,
``Spinors and Space-Time. Vol. 1: Two-Spinor Calculus and Relativistic Fields,''
Cambridge University Press, Cambridge Monographs on Mathematical Physics (1984) 458 pp.

\bibitem{Capovilla:1991kp}
R.~Capovilla, J.~Dell, T.~Jacobson and L.~J.~Mason,
Class.\ Quant.\ Grav.\ \textbf{8}, 41 (1991)
doi:10.1088/0264-9381/8/1/009

\bibitem{MacDowell:1977jt}
S.~W.~MacDowell and F.~Mansouri,
Phys. Rev. Lett. \textbf{38}, 739 (1977)
[erratum: Phys. Rev. Lett. \textbf{38}, 1376 (1977)]
doi:10.1103/PhysRevLett.38.739

\bibitem{Essig:2013lka}
R.~Essig, J.~A.~Jaros, W.~Wester, P.~Hansson Adrian, S.~Andreas, T.~Averett, O.~Baker, B.~Batell, M.~Battaglieri and J.~Beacham, \textit{et al.}
[arXiv:1311.0029 [hep-ph]].


\bibitem{Contaldi:2008iw}
C.~R.~Contaldi, J.~Magueijo and L.~Smolin,
Phys.\ Rev.\ Lett.\ \textbf{101}, 141101 (2008)
doi:10.1103/PhysRevLett.101.141101
[arXiv:0806.3082 [astro-ph]].

\bibitem{Saito:0705.3701}
S.~Saito, K.~Ichiki and A.~Taruya,
JCAP \textbf{09} (2007) 002,
doi:10.1088/1475-7516/2007/09/002
[arXiv:0705.3701 [astro-ph]]. 

\bibitem{Lue:1999rpc}
A.~Lue, L.~Wang and M.~Kamionkowski,
Phys.\ Rev.\ Lett.\ \textbf{83}, 1506 (1999)
doi:10.1103/PhysRevLett.83.1506

\bibitem{Rovelli:2004tv}
C.~Rovelli,
``Quantum Gravity,''
Cambridge University Press, Cambridge (2004).

\bibitem{Alexander:2011qfa}
S.~Alexander, A.~Marciano and R.~A.~Tacchi,
Phys.\ Lett.\ B \textbf{716}, 330 (2012)
doi:10.1016/j.physletb.2012.08.052
[arXiv:1105.3480 [gr-qc]].

\bibitem{LISA:2024hlh}
M.~Colpi \textit{et al.} [LISA],
[arXiv:2402.07571 [astro-ph.CO]].

\bibitem{Caprini:2018mtu}
C.~Caprini and D.~G.~Figueroa,
Class. Quant. Grav. \textbf{35} (2018) no.16, 163001
doi:10.1088/1361-6382/aac608
[arXiv:1801.04268 [astro-ph.CO]].

\bibitem{followup}
S. Alexander, B. Alexandre, J Magueijo and M Pezzelle, in preparation. 

\bibitem{Borissova:2022clg}
J.~N.~Borissova and B.~Dittrich,
Class. Quant. Grav. \textbf{40} (2023) no.10, 105006
doi:10.1088/1361-6382/accbfb
[arXiv:2207.03307 [gr-qc]].

\bibitem{Borissova:2023yxs}
J.~N.~Borissova, B.~Dittrich and K.~Krasnov,
Phys. Rev. D \textbf{109} (2024) no.12, 124035
doi:10.1103/PhysRevD.109.124035
[arXiv:2312.13935 [gr-qc]].







\end{thebibliography}
\end{document}